\documentclass[letter]{aa}
\usepackage{epsfig}
\usepackage{natbib}
\bibpunct{(}{)}{;}{a}{}{,}
\usepackage{graphicx}
\usepackage{txfonts}
\begin{document}
\title{Kappa distribution and hard X-ray emission of solar flares}
\titlerunning{}
\author{J. Ka\v{s}parov\'{a}
        \and
        M. Karlick\'{y}
       }
\institute{Astronomick\'{y} \'{u}stav AV \v{C}R, v.v.i., Fri\v{c}ova 298, 
    251 65 Ond\v{r}ejov, Czech Republic\\
             \email{kasparov@asu.cas.cz}}
\abstract 
{} 
{We investigate whether the so-called kappa distribution, often
used to fit electron distributions detected in-situ in the solar wind, 
can  describe electrons producing the hard X-ray emission in solar flares.} 
{Using Ramaty High Energy Solar Spectroscopic imager (RHESSI) flare
data we fit spatially- and feature-integrated spectra, assuming a kappa
distribution for the mean electron flux spectrum.} 
{We show that a single kappa
distribution generally cannot describe spatially integrated X-ray emission
composed of both footpoint and coronal sources. In contrast, the kappa
distribution is consistent with mean electron spectra producing hard X-ray
emission in some coronal sources.} 
{} 

 \maketitle
\section{Introduction}
In solar flares, electrons are very efficiently accelerated, as shown by
non-thermal X-ray and radio emission 
\citep{ta88,1993ASSL..184.....B,ka97,2007LNP...725.....K}. 
A widely used method to determine the electron distribution functions in solar flares is the
forward-fitting of X-ray spectra in which  
the low-energy part usually is described by isothermal bremsstrahlung while for higher energies,
electron power-law distributions with a low-energy cutoff are assumed \citep[e.g.][]{2003ApJ...595L..97H}.
On the other hand, energetic electrons observed in-situ in space have a distribution
with an enhanced number of particles in the high-energy tail. These electrons
are commonly modelled by kappa distributions 
\citep{1989BAICz..40..175V,1995PhPl....2.2098M,1997A&A...324..725M}. 
This leads us to the idea that the kappa
distribution is a natural consequence of underlying physical processes. In
a series of papers by 
\citet{2006JGRA..11109106Y,2006JGRA..11109107R,2007PhPl...14j0701R}
it was shown analytically and numerically that the
kappa distribution is the byproduct of beam-plasma interaction which first
leads to the excitation of Langmuir waves followed by turbulent mode-coupling
processes. Among the various wave-particle and wave-wave interaction processes,
it was proposed that the ``collisionality'', defined via the number of particles
per Debye sphere, plays a crucial role. 
Meanwhile, it appeared that the
problem of the kappa distribution has an even deeper physical meaning.
\citet{1988JSP....52..479T,2002Ap&SS.282..573L}
have found that the kappa
distribution is the thermodynamic equilibrium solution within so-called
nonextensive thermodynamics with a generalised entropy definition.

Although in the interpretation of the flare X-ray spectra the kappa distribution
has not been considered up to now, it is evident that it has
some attractive properties: a smooth transition between thermal and 
high-energy parts of the X-ray emission producing electrons and no need for any low-energy cutoff. 
Indeed, the recent statistical study by \citet{2008SoPh..252..139K} using a regularised inversion  
indicates that if low-energy cutoffs exist in mean electron spectra, their energies
are likely rather low (below $12$~keV).
Furthermore, taking into account  velocity diffusion
and the slowing-down rate of electrons, \citet{2005A&A...438.1107G} applied a steady-state solution
of the Fokker-Planck equation to RHESSI spectra,
thus consistently treating thermal and fast electrons. Finally, we note that
the kappa distribution is also considered to be connected to the enhanced EUV line emission
observed during solar flares \citep{2005ESASP.600E.120D}.

RHESSI \citep{li02} not only provides high resolution spatially integrated spectra but
it is possible to obtain feature integrated spectra, i.e. spectra from
individual sources, using imaging spectroscopy techniques \citep{2006A&A...456..751B}.

The aim of this paper is to examine whether flare X-ray spatially integrated
spectra or spectra of some particular type of source are consistent with
a kappa distribution of the electrons producing the emission. We use the
forward-fit approach and fit the spectra assuming a kappa distribution of
electrons in the thin-target model.

Properties of the kappa distribution and expressions for the corresponding
bremsstrahlung thin-target emission are presented in Section~\ref{kappa}. The data
analysis is described in Section~\ref{analysis}. In Section~\ref{results} we
show which types of flare spectra and of which sources are consistent with the kappa
distribution of electrons. Our results are summarised in Section~\ref{concl}.

\section{Kappa distribution}\label{kappa}
The kappa distribution $f_\kappa(E)$ represents a generalisation of a Maxwellian
distribution with an enhanced high-energy tail \citep{1991PhFlB...3.1835S}. It
can be written as
\begin{equation}\label{eq:kappa}
f_\kappa(E) = A_\kappa {2 \sqrt{E}\over \sqrt{\pi (k_\mathrm{B}T_\kappa)^3}} \left(1 + {E \over 
(\kappa - 1.5) k_\mathrm{B}T_\kappa}\right)^{-(\kappa + 1)}\,,
\end{equation}
where $A_\kappa=\Gamma(\kappa+1)/[\Gamma(\kappa-0.5)(\kappa-1.5)^{1.5}]$ so that $\int f_\kappa(E)\mathrm{d}E=1$,
$T_\kappa$ is ``temperature'', $k_\mathrm{B}$ is the Boltzmann constant, and $E$ denotes electron kinetic energy. 
At large $E$ and for low $\kappa$,
the distribution is a power-law; in the limit $\kappa\to\infty$ it approaches a Maxwellian distribution.
\subsection{Thin-target X-ray emission}
Assuming that superthermal electrons are described by the distribution
(\ref{eq:kappa}) and the mean electron density $N_\kappa$ in optically thin
plasma volume $V$, the thin-target bremsstrahlung photon spectrum $I(\epsilon)\
[\mathrm{photons}\ \mathrm{cm}^{-2}\ \mathrm{s}^{-1}$~per unit $\epsilon]$ has
the form \citep{br03}
\begin{equation}\label{eq:thin_kappa}
I(\epsilon) = \frac{\overline{n_\mathrm{p}}VN_\kappa}{4\pi R^2} \int_\epsilon^\infty f_\kappa(E)\varv(E)Q(\epsilon,E)\mathrm{d}E
\quad\ \overline{n_\mathrm{p}} = \int_V n_\mathrm{p}\mathrm{d}V / V\,,
\end{equation}
where $Q(\epsilon,E)$ is the bremsstrahlung cross-section differential in photon energy $\epsilon$, $\varv$ denotes velocity,
and $R$ is the distance to the emission source, e.g. 1~AU.
Note that contrary to the widely used power-law forms for electron distributions producing solar flare bremsstrahlung emission,
there is no need to introduce any (often arbitrary) low-energy cutoff. Furthermore, the kappa distribution also describes
the low-energy part of the X-ray emission that is usually fitted by a thermal component.
\section{Data analysis}\label{analysis}
\begin{figure}
\centerline{
\resizebox{0.5\hsize}{!}{\includegraphics{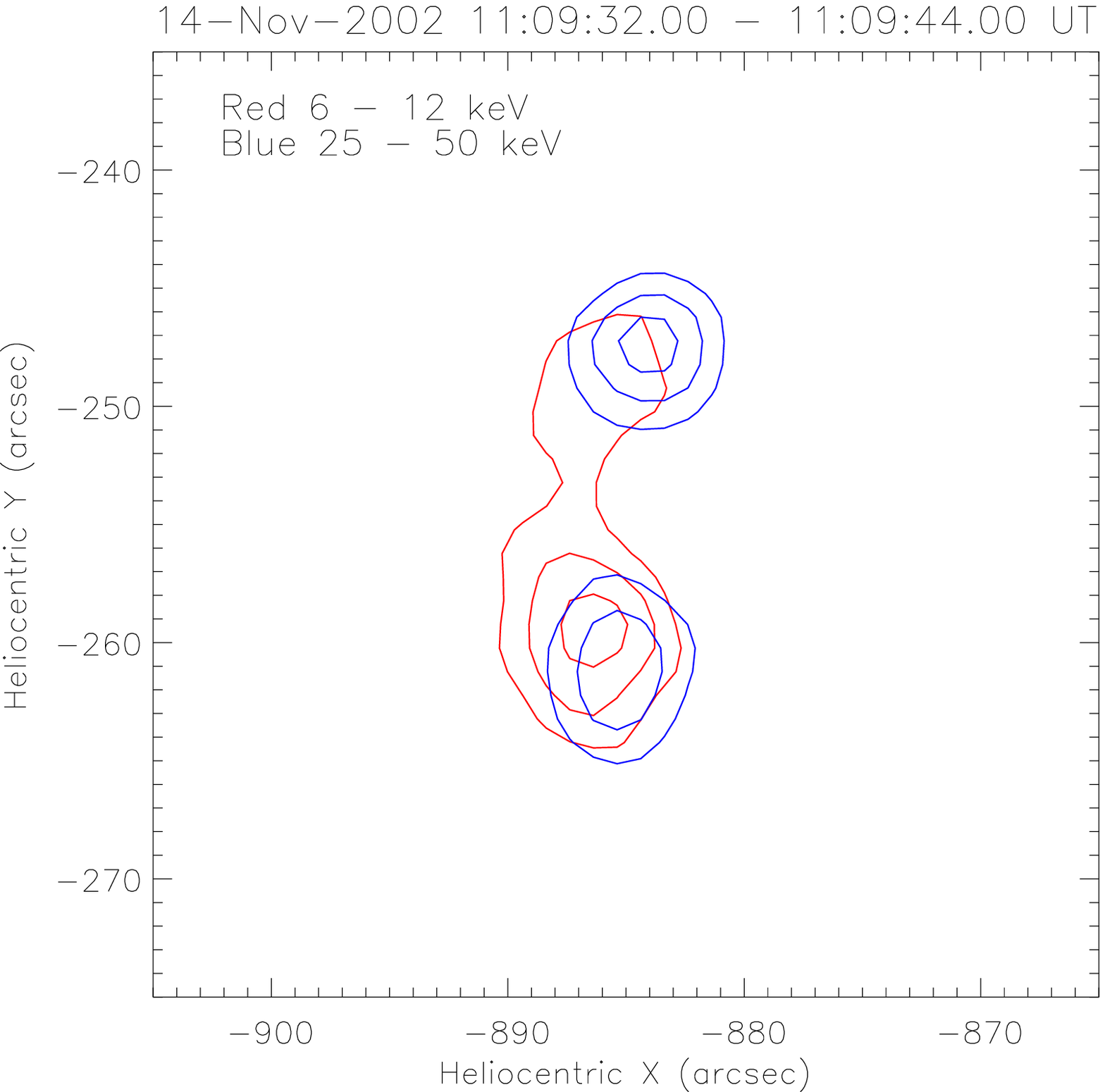}}
\resizebox{0.5\hsize}{!}{\includegraphics{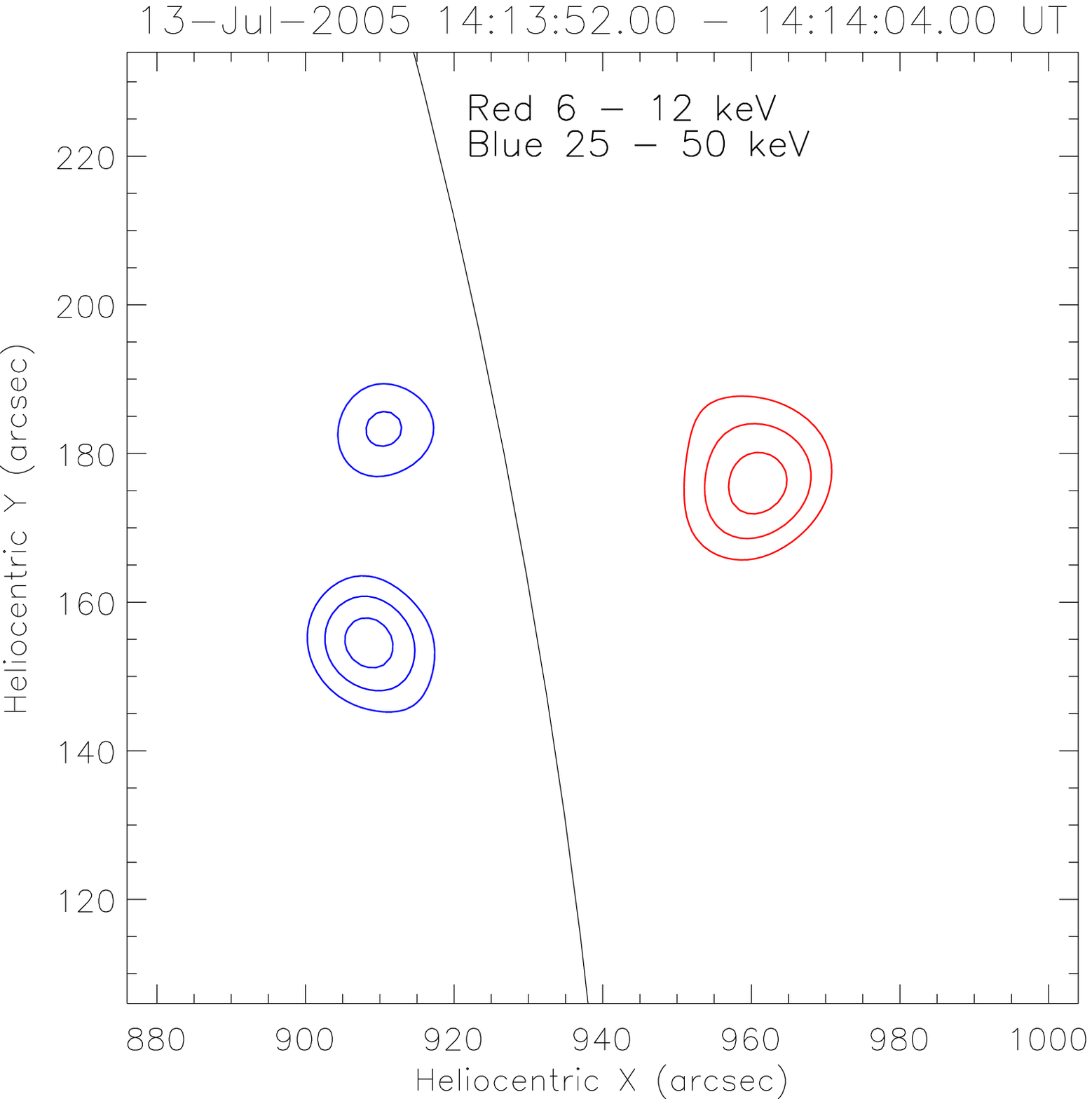}}
}
\vskip2mm
\centerline{
\resizebox{0.5\hsize}{!}{\includegraphics{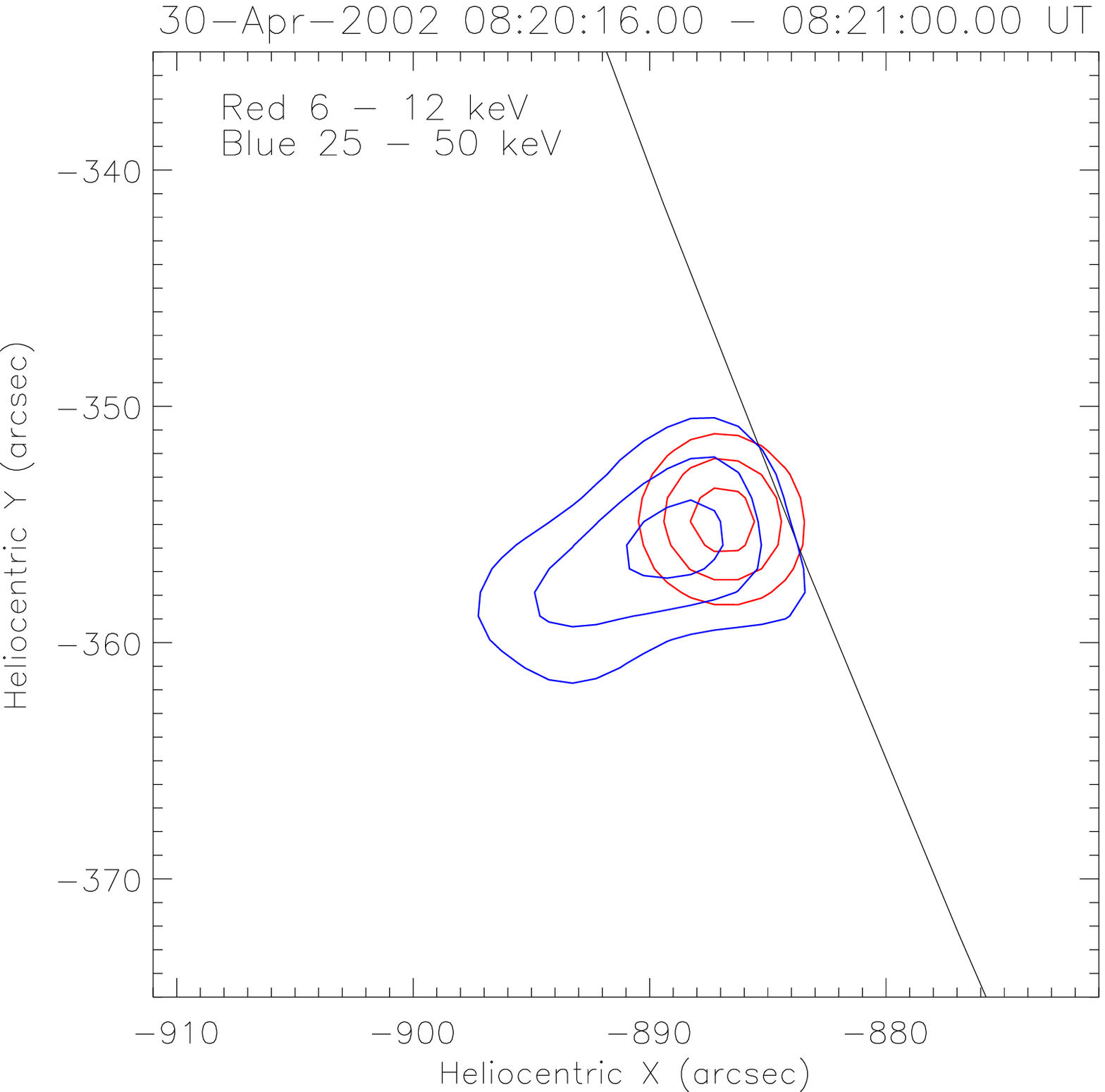}}
\resizebox{0.5\hsize}{!}{\includegraphics{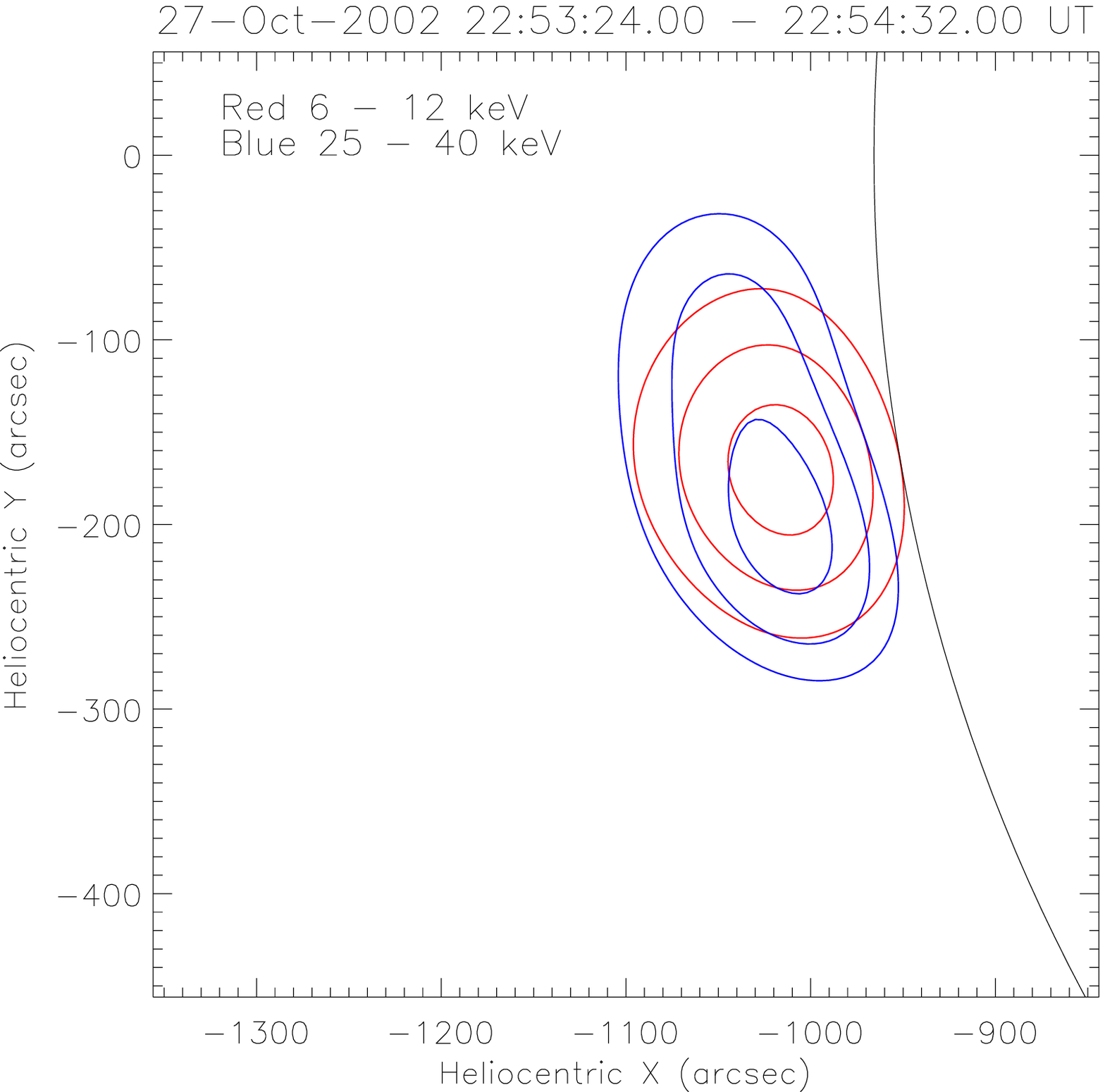}}
}
\caption{Examples of events imaged at low (typically 6~-~12~keV) and high (above 25~keV) energies.}
\label{fig:images}
\end{figure}

In order to explore a range of source types, we considered flares of a
significant thermal component as well as flares with reported non-thermal
behaviour at low energies ($\sim10$~keV) -- the so-called early impulsive flares
studied  by \citet{2007ApJ...670..862S}. Further, events were required to have
well separated sources to obtain spectra of individual sources using imaging
spectroscopy without contamination by other sources. We also analysed partially
occulted flares since their observation could provide us with high-energy
resolution spectra of a single source. To achieve that we selected  partially
occulted events with only one source which was nearly cospatial at thermal
(6~-~12~keV) and higher (typically above 25~keV) energies. Finally, events with
a corrected livetime below 90\% were discarded to avoid pile-up issues.
Figure~\ref{fig:images} shows examples of analysed disc events with several
separated sources ({\it top}) and partially occulted events ({\it bottom}).

Spatially integrated spectra were accumulated during the event peak  (typically tens of seconds) at energies above 25~keV.
Spectra of individual sources were obtained following the imaging spectroscopy approach of \citet{2006A&A...456..751B}.
We used the CLEAN algorithm \citep{2002SoPh..210...61H} to reconstruct images in an interval of several RHESSI spin periods
covering the time interval for the spatially integrated spectra. Using
OSPEX\footnote{http://hesperia.gsfc.nasa.gov/ssw/packages/spex/doc/\\ospex\_explanation.htm}, appropriate
regions of interest were chosen and feature-integrated spectra, i.e. from a footpoint or a coronal source, were obtained.
\subsection{Spectral fitting}\label{sec:fitting}
Spatially and feature-integrated spectra were forward-fitted within OSPEX
assuming the following two distinct fitting models: 
\begin{enumerate}
\item Isothermal component + thin-target component of an electron power-law distribution, i.e. the standard approach.
The fitted parameters are temperature $T$ and emission measure $EM$ of the thermal plasma and parameters
of the power-law, non-thermal, component, i.e. low-energy cutoff $E_\mathrm{c}$, spectral index $\delta_1$,
normalisation factor $\bar{n}V\overline{F}$, and
optionally break energy $E_\mathrm{b}$ and spectral index $\delta_2$ above the break energy in the case of a double power-law.
\item Thin-target emission of an electron kappa distribution (Eq.~(\ref{eq:thin_kappa})) + two Gaussians to account for iron-line complexes
at 6.7 and 8~keV. 
(Note that the contributions of the iron-line complexes are included in the isothermal component in the standard fitting approach.)
The fitted parameters are $T_\kappa$, index $\kappa$, normalisation $\overline{n_\mathrm{p}}VN_\kappa$,
integrated intensities of two Gaussian lines and optionally their centroids $E_1$, $E_2$.
\end{enumerate}
A fit was considered acceptable if the value of reduced $\chi^2$, $\chi^2_\nu$, was close to 1 and normalised residuals did not show
significant clustering at a particular energy.

The thin-target formula for an electron kappa distribution, Eq.~(\ref{eq:thin_kappa}), has been incorporated into the SSW tree and can
be accessed as  OSPEX fitting function \verb(thin_kappa.pro( .
\section{Consistency of the kappa distribution with flare X-ray spectra}\label{results}
Applying two fitting models described in Section~\ref{sec:fitting} to spectra of several flares, we have found
that the spatially integrated spectra of disc events cannot  be fitted with the electron kappa distribution.
The best fits using the kappa distribution have large value of $\chi^2_\nu$ and large residuals, generally around 10~keV.
On the contrary, those spectra are easily fitted by the standard approach, i.e. an isothermal component
plus an electron power-law spectrum. Such spectra are mostly a combination of emission coming from
several sources (foopoints, coronal or loop source -- see Fig.~\ref{fig:images}, top row) 
of distinct plasma properties, therefore it is difficult to describe
the whole spectrum by just one distribution. On the other hand, using the combination of an isothermal
and a power-law component allows us to adjust parameters of these functions in such a way that a satisfactory fit usually can be
obtained. The low-energy cutoff helps describe the smooth transition between the thermal and
non-thermal components independently of other fit parameters.
Conversely, the shape of the kappa function is given only by two parameters: $T_\kappa$ and $\kappa$.

\begin{figure}
\resizebox{\hsize}{!}{\includegraphics{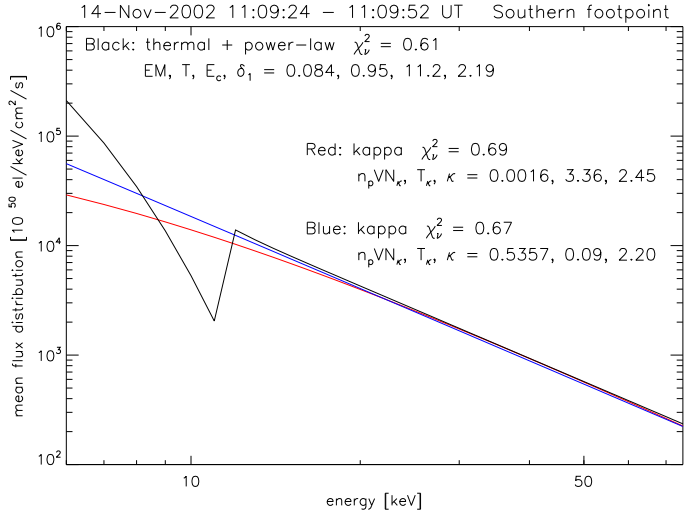}}
\caption{Comparison of electron distributions obtained by fitting emission from the southern footpoint of 
the 14-Nov-2002 flare (see also Fig.~\ref{fig:images}).
Black lines denotes the isothermal plus power-law component fit, red and blue lines correspond to kappa distribution fits. Due to large
uncertainties in the data, similarly good fits in terms of $\chi^2_\nu$, but differing at energies below $\sim 10$~keV,  
can be obtained.}
\label{fig:14nov02footpointfbar}
\end{figure}

Next, we examined the feature integrated spectra obtained by the imaging
spectroscopy approach. Spectra related to footpoints can be fitted by both
models; the best fits are similar in terms of $\chi^2_\nu$ and residuals. The
footpoint emission at energies above 10~keV is approximated well by the kappa
distribution; $\kappa$ is generally below 4, therefore the resulting
distribution resembles the power-law used in the standard approach. However,
parameters $T_\kappa$ and $\overline{n_\mathrm{p}}VN_\kappa$ are not constrained in
the model with the kappa distribution. Equally good fits can be obtained with
values of $T_\kappa$ and $\overline{n_\mathrm{p}}VN_\kappa$ differing by an order of
magnitude -- compare the resulting electron distributions and values of fitted parameters in
Fig.~\ref{fig:14nov02footpointfbar}. This is probably related to the fact
that footpoint emission at low energies is reconstructed with high
uncertainties. Therefore, we conclude that there is no advantage in using a kappa
distribution for the footpoint spectra.

\begin{figure}
\resizebox{\hsize}{!}{\includegraphics{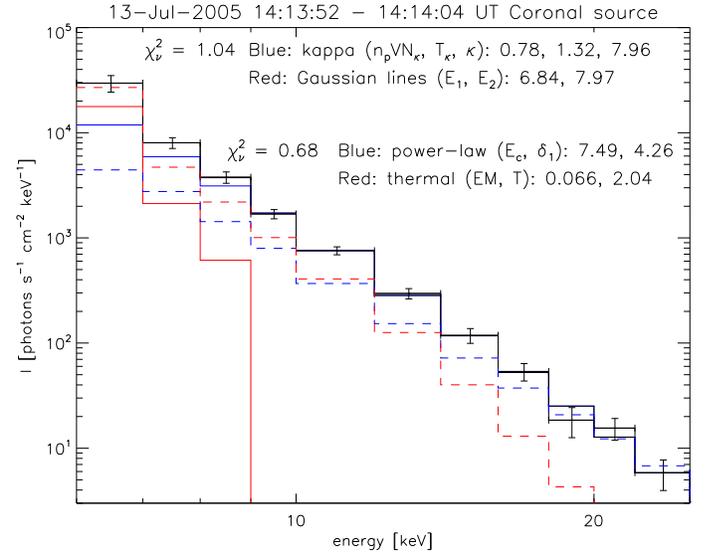}}
\caption{Spectrum of the coronal source of the 13-Jul-2005 flare (see also Fig.~\ref{fig:images}) obtained from CLEAN images.
Data are plotted as crosses; their horizontal sizes correspond to the widths of energy bins, vertical sizes represent flux uncertainties. 
Coloured solid lines denote the components of the kappa plus lines fit, dashed lines correspond to the components of
the isothermal plus power-law fit. The thick black line shows the resulting fit using the kappa distribution.
}
\label{fig:13jul05fit}
\end{figure}

As concerns coronal sources of disc events, the 13-Jul-2005 flare (see Fig.~\ref{fig:images}, top right)
coronal emission can be well fitted by both spectral models. Both models provide best fits with similar
residuals and $\chi^2_\nu$ close to 1 (see Fig.~\ref{fig:13jul05fit}). Moreover, all kappa parameters are well constrained.

\begin{figure}
\resizebox{0.95\hsize}{!}{\includegraphics{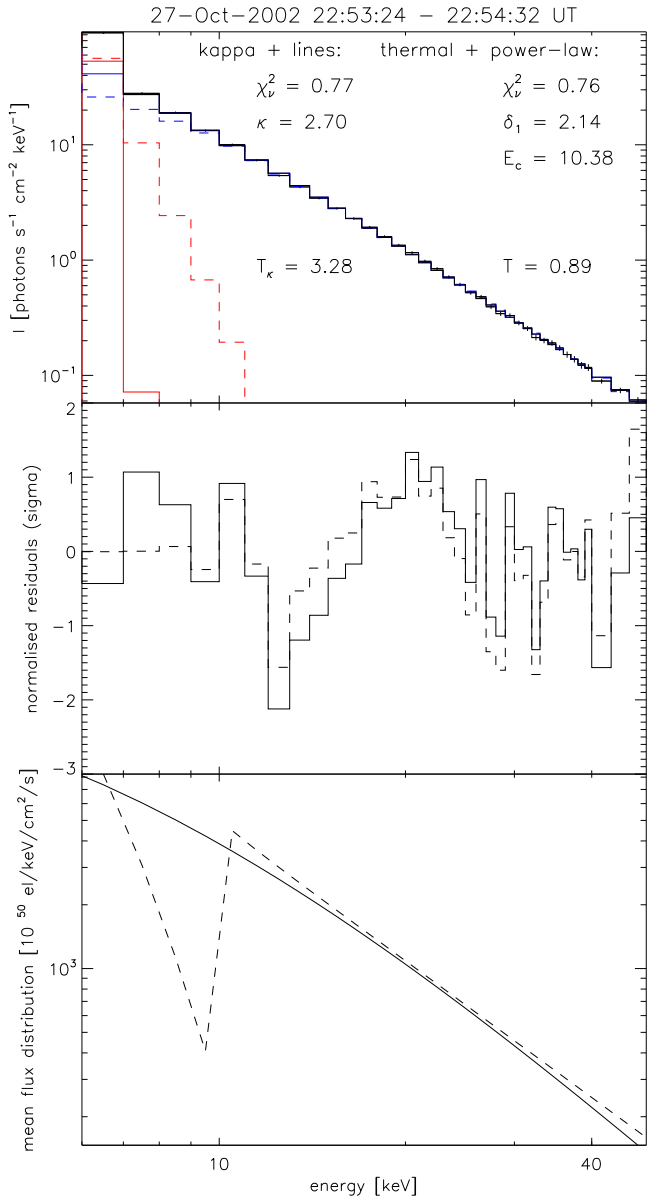}}
\caption{
Comparison of the kappa distribution fit (solid) and the isothermal plus power-law fit (dashed)
to the spatially integrated spectrum of the partially occulted flare on 27-Oct-2002 (see also Fig.~\ref{fig:images}).
Colour coding is the same as in Fig.~\ref{fig:13jul05fit}. 
{\it Top:} observed photon spectrum (crosses -- horizontal and vertical sizes represent widths of energy bins and flux uncertainties, respectively) 
and fits (histograms), 
{\it middle:} normalised residuals of the fits, {\it bottom:} resulting
electron distributions.
}
\label{fig:27oct02fit}
\end{figure}

Having obtained a reliable fit using the kappa function to a coronal source spectrum of a disc event reconstructed from CLEAN images,
we examined high-energy resolution spatially integrated spectra of partially occulted flares.
Such spectra are generally obtained with much finer energy resolution and better signal-to-noise ratio than
the data from imaging spectroscopy (compare uncertainties on the data points in Figs.~\ref{fig:13jul05fit} and \ref{fig:27oct02fit}).
Partially occulted events were recently studied by
\citet{2008ApJ...673.1181K} and we analysed several events listed in that paper.
A representative example is the 30-Apr-2002 event.
Its spectrum cannot be described well by the model using a kappa function (the best fit shows large residuals at energies around 10~keV),
whereas the standard model matches the spectrum reasonably well. This event shows nearly cospatial sources at 6~-~12 and 12~-~25~keV ranges,
yet their shapes suggest different structures depending on energy  -- see Fig.~\ref{fig:images}, bottom left.

On the other hand, the spatially integrated spectrum of the extended coronal source of 27-Oct-2002 flare seen by RHESSI above the limb
(see Fig.~\ref{fig:images}, bottom right)
can be fitted well by the kappa function up to 50~keV.
Again, the standard model provides a similarly good fit both in terms of residuals and $\chi^2_\nu$, but contrary to the
kappa function model a low-energy cutoff must be introduced. The best fit using the kappa function is shown in Fig.~\ref{fig:27oct02fit}.

Now the question arises of why the kappa distribution fits the spectrum of the
October 27, 2002 coronal source much more closely than that of April 30, 2002.
\citet{2007ApJ...669L..49K} show that the October 27, 2002 source 
was observed at much higher altitudes (0.3 $R_{\sun}$) than 
the April 30, 2002 event. We think that at lower altitudes the X-ray source structure can be
more complex (e.g. several different X-ray sources along the line of sight)
than for sources at higher heights and thus the fit of the kappa
distribution is not successful due to the limited number of free parameters.
\section{Conclusions}\label{concl}
We have applied a kappa distribution of electrons to RHESSI hard X-ray flare spectra. We have shown that spatially integrated spectra
of disc events, generally composed of contributions from coronal and footpoint emissions, cannot be well described
as the thin-target emission of an electron kappa distribution. On the contrary, the standard approach assuming an isothermal
and a power-law component results in a better fit in terms of residuals and reduced $\chi^2$.
This could be due to the fact that those different spectral components come from regions of distinct plasma properties
and therefore it is difficult to describe the spectral shape by two parameters, $T_\kappa$ and $\kappa$ only.

Further, from the spectra of individual sources obtained by means of imaging spectroscopy,
we have found that both footpoint and coronal emissions could be fitted using the electron kappa function. However,
in the case of the footpoints, the fit is ambiguous at low energies owing to high uncertainties of data
in this energy range. Therefore, we cannot conclude whether the kappa distribution is appropriate for the footpoint emission.

In constrast, the results both from imaging spectroscopy and high-energy resolution spectra
of partially occulted flares suggest that some coronal sources are consistent
with the thin-target emission of the electron kappa distribution.

Summarising the results of all analysed flares, it looks as if the decrease of
X-ray source complexity increases the relevance of spectral fitting by the
kappa distribution, especially for coronal X-ray sources. This is in
agreement with the numerical models of \citet{2006JGRA..11109106Y}
and \citet{2006JGRA..11109107R} who consider
particle-wave interactions in plasma and thus describe processes appropriate
for coronal conditions where Coulomb collisions do not dominate.

Our results do not rule out the widely used interpretation of hard X-ray spectra being produced by thermal and power-law components, yet
they show that other distributions, such as the kappa distribution, are consistent with the RHESSI data as well.
In our view, the advantage of using the kappa distribution lies mainly in the facts that
the kappa distribution is a natural consequence of physical processes in space plasma, that
such a distribution is applicable to in-situ detected electrons, and no low-energy cutoff in the mean electron
spectrum has to be considered.

We note that the low-energy part of hard X-ray spectra is important to constrain the best fit. For future studies we propose to combine
hard X-ray data with a detailed
analysis of soft X-ray continuum and lines from other instruments such as RESIK \citep{2005SoPh..226...45S}
or the recently launched SphinX on board Coronas-Photon \citep{2008JApA...29..339S}.
Finally, on the basis of nonextensive thermostatistics, \citet{2004ApJ...604..469L} proposed a so-called double-kappa distribution
and applied it to observed interplanetary electron distributions.
Future RHESSI data analyses may use this idea as an alternative approach to the single kappa distribution used here.

\begin{acknowledgements}
We are grateful to E. Dzif\v{c}\'{a}kov\'{a} for her valuable remarks concerning the kappa
distribution. This work was supported by grants 205/06/P135,
205/09/1705 (GA \v{C}R), IAA300030701 (GA AS \v{C}R), 
and the research project AV0Z10030501 (Astronomick\'{y} \'{u}stav).
\end{acknowledgements}
\bibliographystyle{aa}
\bibliography{references}
\end{document}